\documentclass[english,twocolumn,aps,prb,footinbib,amssymb,floatfix,superscriptaddress,longbibliography,showpacs]{revtex4-1}
\usepackage[T1]{fontenc}
\usepackage[latin9]{inputenc}
\usepackage{textcomp,amsfonts,amstext,amsmath,graphicx,color,hyperref,siunitx,babel,times,float}

\makeatletter

\providecommand{\tabularnewline}{\\}

\usepackage{CJK} % will be needed for Ming-Hao's name

\makeatother

\begin{document}
\begin{CJK}{UTF8}{bmsi}
	
\title{Probing Spin Helical Surface States in Topological HgTe Nanowires}
\begin{abstract}
... 
\end{abstract}

\author{J.~Ziegler}

\address{Institut f{\"u}r Experimentelle und Angewandte Physik, Universit{\"a}t Regensburg,
93053 Regensburg, Germany}

\author{R.~Kozlovsky}

\author{C.~Gorini}

\address{Institut f{\"u}r Theoretische Physik, Universit{\"a}t Regensburg, 
93053 Regensburg, Germany}

\author{M.-H.~Liu ({\CJKfamily{bsmi}劉明豪})}
%\author{M.-H.~Liu}
\address{Institut f{\"u}r Theoretische Physik, Universit{\"a}t Regensburg, 
93053 Regensburg, Germany}
\address{Department of Physics, National Cheng Kung University, Tainan 70101, Taiwan}

\author{S.~Weish{\"a}upl}

\author{H.~Maier}

\author{R.~Fischer}

\address{Institut f{\"u}r Experimentelle und Angewandte Physik, Universit{\"a}t Regensburg,
93053 Regensburg, Germany}

\author{D.~A. Kozlov}

\author{Z.~D. Kvon}

\address{A.V. Rzhanov Institute for Semiconductor Physics, Novosibirsk, Russia}

\address{Novosibirsk State University, Russia}

\author{N.~Mikhailov}

\author{S.~A. Dvoretsky}

\address{A.V. Rzhanov Institute for Semiconductor Physics, Novosibirsk, Russia}

\author{K.~Richter}

\address{Institut f{\"u}r Theoretische Physik, Universit{\"a}t Regensburg, 
93053 Regensburg, Germany}

\author{D.~Weiss}

\address{Institut f{\"u}r Experimentelle und Angewandte Physik, Universit{\"a}t Regensburg,
93053 Regensburg, Germany}

\date{\today}

\begin{abstract}
%%%
Nanowires with helical surface states represent key prerequisites for observing and exploiting phase-coherent topological conductance phenomena, such as spin-momentum locked quantum transport or topological superconductivity. 
%%%
We demonstrate in a joint experimental and theoretical study that gated nanowires fabricated from high-mobility strained HgTe, known as a bulk topological insulator, indeed preserve the topological nature of the surface states, that moreover extend phase-coherently across the entire wire geometry.
%%%
The phase-coherence lengths are enhanced up to 5 $\mu$m when tuning the wires into the
bulk gap, so as to single out topological transport.
%%%
The nanowires exhibit distinct conductance oscillations, both as a function of the flux due to an axial magnetic field, and of a gate voltage.
%%%
The observed $h/e$-periodic Aharonov-Bohm-type modulations indicate surface-mediated quasi-ballistic transport.
%%%
Furthermore, an in-depth analysis of the scaling of the observed gate-dependent conductance oscillations reveals the topological nature of these surface states.
%%%
To this end we combined numerical tight-binding calculations of the quantum magneto-conductance with simulations of the electrostatics, accounting for the gate-induced inhomogenous charge carrier densities around the wires.
%%%
We find that helical transport prevails even for strongly inhomogeneous gating and is governed by flux-sensitive high-angular momentum surface states that extend around the entire wire circumference.
%%%
\end{abstract}


 \maketitle
\end{CJK}

\section{Introduction}

Three-dimensional topological insulators (3DTIs) are a particular class of bulk insulators hosting time reversal symmetry-protected metallic surface states. The latter are helical, {\em i.e.}~characterized by (pseudo)spin-momentum locking, and described by low-energy effective Dirac-type models \cite{Hasan2010}. In nanowires based on 3DTI materials
such locking heavily affects the one-dimensional (1D) subband spectrum and, if combined with superconductivity, is a basic ingredient for the realization of Majorana modes \cite{Hasan2010, BardarsonMoore2013}. Moreover, from a general quantum transport perspective, 3DTI nanowires provide a particularly rich playground due to the interplay between topological properties and effects arising from phase coherence. The fact that the conducting states are ``wrapped'' around an
insulating bulk, in conjunction with their helical nature, leads to various interesting and geometry-sensitive 
magnetoresistive phenomena \cite{BardarsonMoore2013,Sacksteder2014} that are inaccessible in standard metallic systems, whose bulk and surface contributions cannot in general be singled out.

\begin{figure}[t]
\includegraphics[width=1\columnwidth]{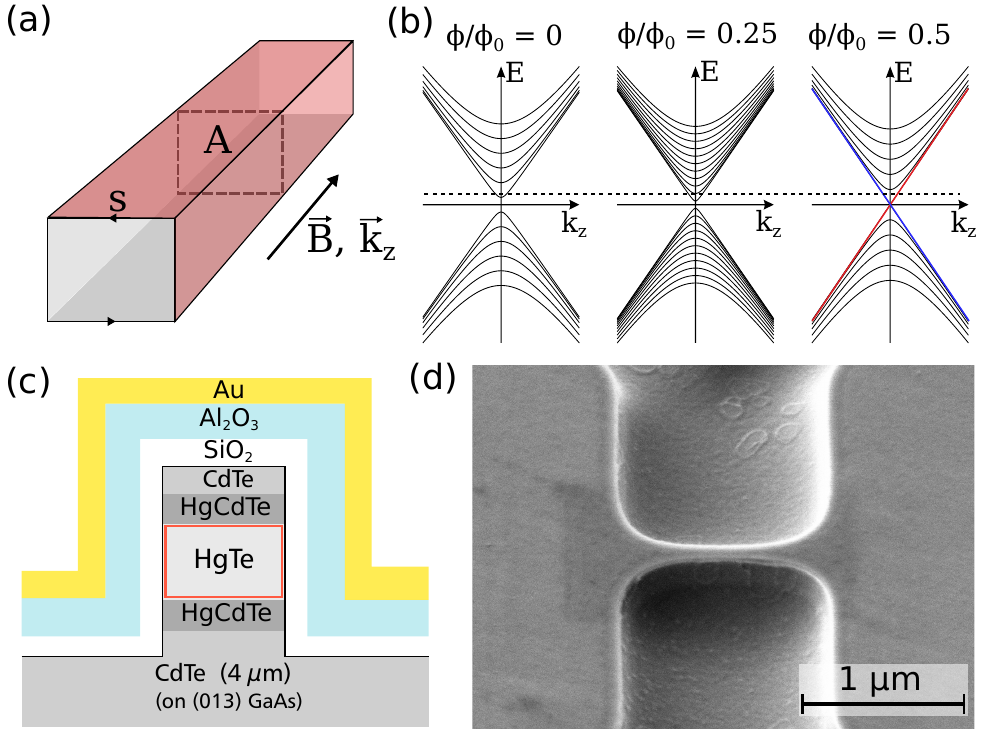} \caption{\label{fig1} HgTe-based nanowire. 
(a) Sketch of a nanowire with cross section $A$ and magnetic field $\vec{B}$ along the wire axis. 
(b) 1D band structures of the surface states of three-dimensional topological insulator nanowires for three different representative magnetic fluxes. 
(c) Schematic cross section of a nanowire employed in experiment. The nanowires are etched from wafers containing an 80~nm strained HgTe film and consist of a 4~\textmu m CdTe base layer grown on (013) GaAs substrate, two 20~nm Hg$_{0.7}$Cd$_{0.3}$Te buffer layers on top and bottom of the HgTe and a 40~nm CdTe cap layer. After wet chemical etching the wire is covered by SiO$_{2}$/Al$_{2}$O$_{3}$ and a metallic top gate. 
(d) SEM micrograph of a representative HgTe TI nanowire, taken at a tilt angle of 50\textdegree{} before deposition of the topgate structure. It has a length of 1.3 \textmu m and a median width of 163 nm.}
\end{figure}

In particular, a 3DTI nanowire in a coaxial magnetic field with magnitude $B$ and associated flux $\phi = AB$, as sketched in Fig.\,\ref{fig1}(a), is expected to show peculiar Aharonov-Bohm type magnetoresistance features. Indeed, oscillations with a period of one flux quantum $\phi_0 = h/e$ (where $h=2\pi\hbar$ is Planck's constant and $e$ the elementary charge)  were observed in early experiments \cite{PengLaiKongEtAl2010,Xiu2011}. According to theory \cite{Ostrovsky2010,BardarsonBrouwerMoore2010,Zhang2010,Rosenberg2010,BardarsonMoore2013} these oscillations reflect the wire's 1D subband structure, 
given by
\begin{equation}
E=\pm\hbar v_{F}\sqrt{k_{z}^{2}+k_{l}^{2}}\quad\text{with}\quad k_{l}=\frac{2\pi}{P}\left(l+\frac{1}{2}-\frac{\phi}{\phi_{0}}\right) \label{eq1}.
\end{equation}

In Eq.~(\ref{eq1}), $v_{F}$ is the Fermi velocity of the surface carriers, 
$k_z$ the coaxial and $k_l$ the transversal wave vector, the latter having the meaning of angular momentum.
The angular momentum quantum number is labeled by $l \in \mathbb{Z}$ and its half-integer shift in $k_l$ is caused by a
curvature induced Berry phase. The resulting energy spectrum is sketched in Fig.\,\ref{fig1}(b) for three characteristic
values of the magnetic flux. 
For $\phi=0$ an energy gap is present (due to the Berry phase) and the 1D subbands are twofold degenerate with respect to angular momentum.
Note, however, that owing to their Dirac-like nature these states are not spin-degenerate. 
For finite flux such as $\phi/\phi_{0}=0.25$ the degeneracies with respect to $k_l$ are lifted. 
For $\phi/\phi_{0}=0.5$, the magnetic flux cancels the Berry phase and the $l=0$ states become gapless, $k_{l=0}=0$.  
This linear gapless state is non-degenerate. More generally, the total number of states is odd, and time-reversal 
symmetry, restored at $\phi/\phi_{0}=0.5$, implies one ``perfectly transmitted mode'' \cite{Ando2002}.
For $\phi/\phi_0 = 1$ the spectrum recovers its $\phi/\phi_0 = 0$ form.
Further increasing the flux leads to analogous cycles: the Berry phase cancellation, and thus the appearance
of a gapless state, takes place for half-integer values of $\phi/\phi_0$, whenever $k_l=0$ for a given $l$,
while for integer values the gapped spectrum for $\phi=0$ in Fig.~\ref{fig1}(b) is recovered.

This ``shifting'' of the 1D subbands with changing flux implies that for a given Fermi level position 
the conductance should be $\phi_0$-periodic. 
Dominating $h/e$ oscillations were indeed observed in a number of experiments \cite{Tian2013,DufouleurVeyratTeichgraeberEtAl2013,Safdar2013,Hong2014,ChoDellabettaZhongEtAl2015,JaureguiPettesRokhinsonEtAl2016,Kim2016}. 
Furthermore, the phase of the Aharonov-Bohm-type oscillations depends on the Fermi level position.  
To be definite, consider $E_{F}$ close to zero: no states are present for $\phi=0$ 
(and corresponding even multiples of $\phi_0/2$), whereas for $\phi/\phi_{0}=0.5$ 
(and corresponding odd multiples of $\phi_0 / 2$) the gapless mode emerges. 
Hence, one expects a conductance minimum (maximum) for integer (half-integer) flux quanta. 
For $E_F$ slightly above zero, e.~g.~at the dashed line in Fig.\,\ref{fig1}(b), the situation is reversed: 
two modes are present for $\phi=0$, and only one for $\phi/\phi_{0}=0.5$. 
Such $\pi$-phase shifts as a function of the Fermi level position, {\em i.~e.}~the gate voltage $V_{g}$, were observed in
Bi$_{1.33}$Sb$_{0.67}$Se$_{3}$ \cite{ChoDellabettaZhongEtAl2015} and Bi$_{2}$Te$_{3}$ \cite{JaureguiPettesRokhinsonEtAl2016}. 
If the Fermi level can be tuned to the Dirac point a conductance minimum
for $\phi/\phi_{0}=0$ and a maximum at $\phi/\phi_{0}=0.5$, as observed in \cite{ChoDellabettaZhongEtAl2015}, 
are signatures of the Berry phase of $\pi$ and thus of spin-helical Dirac states. 
However, in strained HgTe nanowires investigated here, the Dirac point is buried in the valence 
band \cite{Brune2011,Crauste2013,Wu2014} and thus cannot be singled-out and probed on its own.
Without direct access to the latter, the phase switching alone is not an exclusive signature of Dirac states:
the 1D subband spectrum of trivial surface states, and thus the resulting conductance, would also be $\phi_0$-periodic.

Hence the crucial question arises how to distinguish topological from trivial states in 3DTI nanowires with an
inaccessible Dirac point, such as strained HgTe.
This is the central point that we address below, by quantitatively analyzing the conductance oscillation 
periodicity occurring as a function of gate voltage $V_{g}$ at fixed flux. 
The observed oscillations reflect directly the 1D subband structure and its degeneracies.  
This allows us to draw conclusions about the nature of the surface states, trivial ones being spin-degenerate, 
in contrast to spin-helical Dirac states. In doing so, we also address a second issue,
namely the consequences of a varying carrier density around the wire circumference.  
Experiments typically rely on the use of top and/or back gates, which couple differently 
to the top, bottom and side surfaces of the 3DTI wires. The carrier density (or the capacitance) therefore 
becomes a strongly varying function of the circumference coordinate $s$. 
Note also that a certain degree of inhomogeneity is expected even in the absence of gating, 
as a consequence of intrinsic system anisotropies \cite{Silvestrov2012}.
As we show below, an inhomogeneous surface charge distribution modifies significantly the band structure, 
yet leaves the essential physics intact.

\section{Nanowire devices}

\begin{table}
\begin{centering}
\begin{tabular}{|c|c|c|c|c|}
\hline 
device  & $\overline{w}$ (nm)  & $l$ (\textmu m)  & $P$ (nm)  & $\Delta B_{h/e}$ (T)\tabularnewline
\hline 
\hline 
t1  & 302  & 0.99  & 724  & 0.203\tabularnewline
\hline 
t2  & 518  & 2.06  & 1156  & 0.116\tabularnewline
\hline 
t3  & 246  & 2.51  & 613  & 0.249\tabularnewline
\hline 
w1  & 310  & 1.33  & 740  & 0.197\tabularnewline
\hline 
w2  & 163  & 1.33  & 446  & 0.386\tabularnewline
\hline 
w3  & 178  & 1.06  & 476  & 0.351\tabularnewline
\hline 
w4  & 294  & 1.95  & 708  & 0.208\tabularnewline
\hline 
w5  & 287  & 2.97  & 694  & 0.212\tabularnewline
\hline 
\end{tabular}
\par\end{centering}
\caption{\label{tab1} Geometrical parameters of the nanowire structures considered. The height $h$ of all devices is given by the 80~nm thick HgTe layer of the wafers. Median width and length of the structures are denoted by $\overline{w}$ and  $l$, respectively. The circumference $P$ and the expected period for $h/e$-oscillations $\Delta B_{h/e}$ (in Tesla) were calculated assuming the surface state wave functions to be located $\sim5$~nm beneath the bulk surface. }
\end{table}

The investigated TI nanowires were fabricated from (013) oriented, strained $80$~nm thick HgTe thin films grown by
molecular beam epitaxy on (013) oriented GaAs substrates (for details see Ref.\ \cite{Dantscher2015}). A cross section
through the layer sequence is shown in Fig.\,\ref{fig1}(c). Using electron beam lithography and wet chemical etching, nanowires as the one shown in Fig.\,\ref{fig1}(d) were fabricated. For etching we used a Br$_{2}$-based wet etch process to preserve the high charge carrier mobilities of the bulk material. The wet etching did not result in perfectly rectangular
wire cross sections but rather in trapezoidal ones with the narrower side on top. In Fig.\,\ref{fig1}(d) the top width of the wire's central segment was $133$~nm while the bottom width was $193$~nm, resulting in an average (median) width of $163$~nm. For easier modeling below we use a rectangular cross section with area $A$ given by the average wire width times HgTe film thickness, $A=\overline{w}h$. This approximation reduces the circumference $P$ that determines the spacing between angular momenta, $\Delta
k_{l}=2\pi/P$, by $2-3\%$. As discussed below, the resulting effect of this assumption is negligible. After etching the wires were covered with $30$~nm of Si$_{2}$O$_{3}$ using plasma enhanced chemical vapor deposition and $100$~nm Al$_{2}$O$_{3}$ deposited with atomic layer deposition. For gating a metallic top layer consisting of titanium and gold was used. The resulting schematic cross section is depicted in Fig.\,\ref{fig1}(c) and will be used for the electrostatic modeling in Sec.~\ref{sec:electrostatic model}. Ohmic contacts to the wire were formed via soldered indium. The nanowires were fabricated in both a true 4-terminal geometry (devices denoted by w3-w5) as well as a quasi-2-terminal geometry (devices t1-3, w1-2). In the latter case, the nanowire constriction was embedded in a larger Hall bar, where the voltage probes are several \textmu m removed from the device.

\begin{figure}
\includegraphics[width=1\columnwidth]{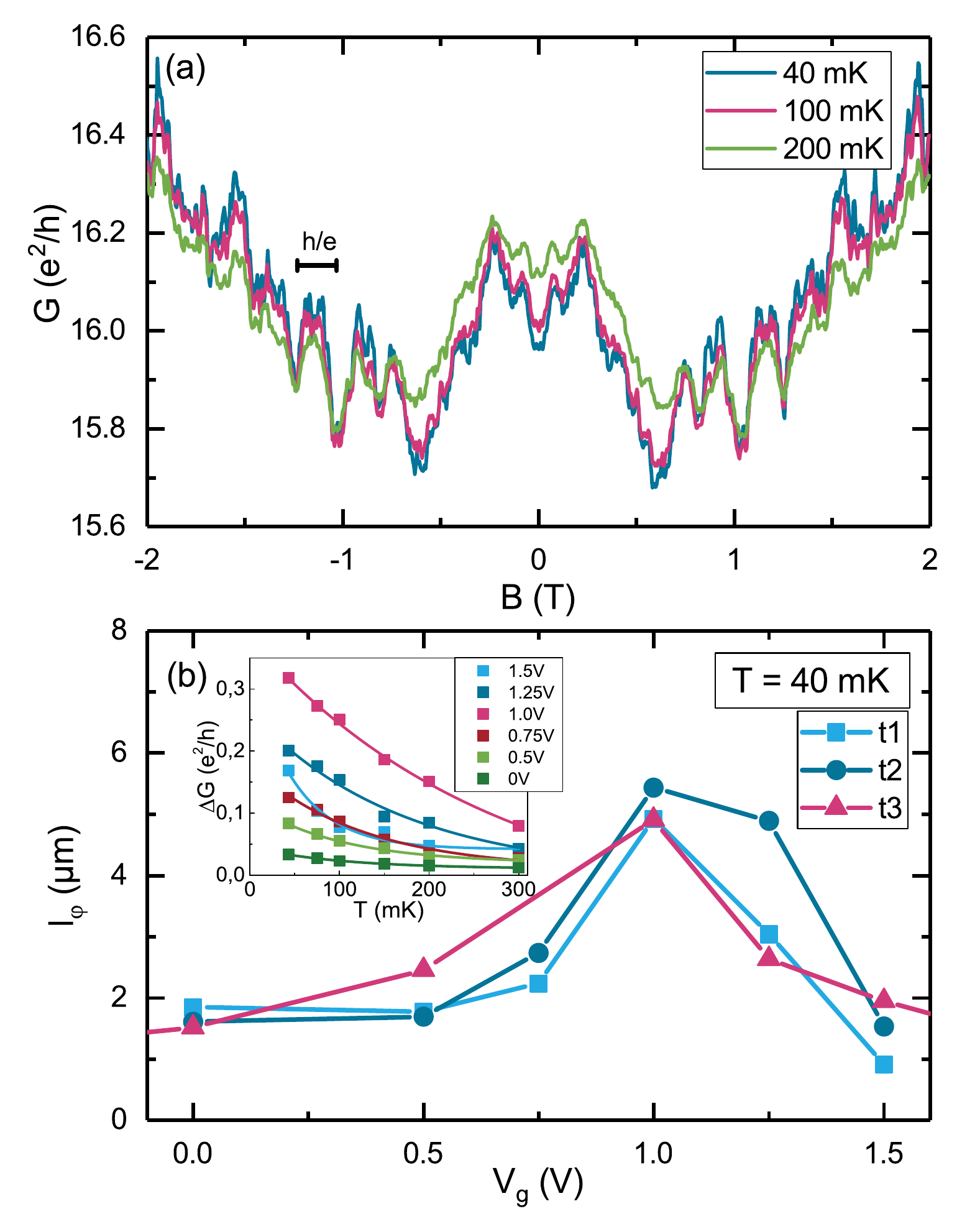} \caption{\label{fig2} (a) Magnetoconductance oscillations of device t1 measured at different temperatures $T$. The traces show $h/e$-periodic oscillations with amplitudes $\Delta G(T)$ decreasing with increasing temperature $T$. The $T$-dependence of the amplitude of sample t1, taken at various gate voltages, is shown in the inset of Panel (b), along with the corresponding fits using Eq.~(\ref{eq2}). The exponential fits allow for the extraction of $l_{\varphi}$, displayed in Panel (b) as a function of gate voltage $V_{g}$ for devices t1-t3.}
\end{figure}

The magnetoconductance of the wires was measured in a dilution refrigerator at temperatures of typically 50~mK and magnetic fields up to 5~T. A rotating sample holder was used for measurements to allow for in- and out-of-plane alignment of the magnetic field. Standard AC lock-in techniques and Femto voltage preamplifiers were used at excitation amplitudes and frequencies of typically $1$~\textendash~$15$~nA and $7$~\textendash ~$13$~Hz, respectively. Additionally, a cold RC-filter was added to suppress noise at the top gate and to improve reproducibility of magnetotransport traces. In the following, we present results from a total of 8 nanowires. Parameters defining their geometries are listed in Table\,\ref{tab1}. Devices t1-t3 were investigated regarding the temperature dependence of the Aharonov-Bohm-type oscillations (see Sec.\,\ref{subsec:AB-oscillations}), while devices w1-w5 were studied with regard to signatures of the subband structure (see Sec.\,\ref{subsec:subbands}).

\section{characterization of conductance oscillations
\label{sec:characterization}}
\subsection{
Magnetoconductance 
\label{subsec:AB-oscillations}
}

\begin{figure}
\includegraphics[width=1\columnwidth]{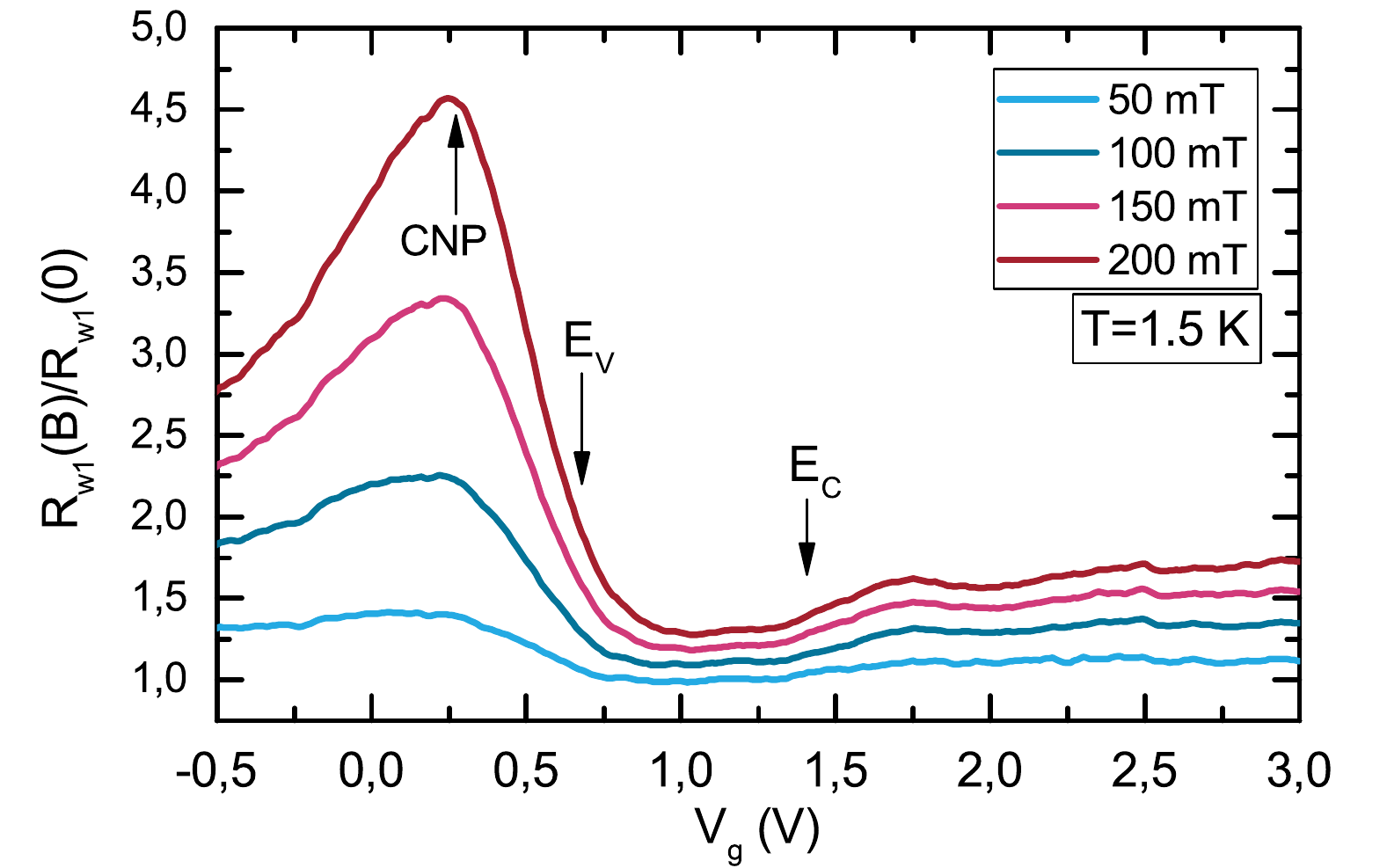} \caption{\label{fig3}
       Normalized longitudinal resistance $R_{w1}(B)/R_{w1}(B=0)$ as a function of gate voltage for nanowire device w1. The upper and lower bounds of the bulk band gap are indicated by arrows and where determined by comparing to analysis done on material with the same wafer stack by Kozlov \textit{et al}. \cite{KozlovKvonOlshanetskyEtAl2014}.
	}
\end{figure}

Figure \ref{fig2}(a) shows the measured two-point conductance $G$ as a function of a magnetic field $B$ applied along the wire axis. The experimental data was taken at temperatures $T$ between 40~mK and $ 300 $~mK. An overall $h/e$ periodicity is clearly visible, as indicated by the horizontal bar in Fig.~\ref{fig2}(a) (and analyzed in more detail below). The fact that $G(B)$ exhibits Aharonov-Bohm-type $h/e$ periodicity instead of $h/(2e)$ behavior, arising from interference between time-reversed paths in the diffusive limit \cite{Aronov1987}, implies that transport along the wires is indeed non-diffusive, {\em i.e.} the elastic mean free path is presumably not much shorter than the wire length and larger
than the wire circumference. However, the additional conductance fluctuations present in $G(B)$ in Fig.~\ref{fig2}(a) indicate residual disorder scattering.

From the temperature dependence we can estimate the phase coherence length $l_{\varphi}$ by using the exponential decay of the amplitude $\Delta G$ of the $h/e$ conductance oscillations for ballistic transport on scales of the perimeter $P$ \cite{Washburn1986},
\begin{equation}
\Delta G\propto \exp\left(-\frac{P}{l_{\varphi}(T)}\right)\,.\label{eq2}
\end{equation}
In Fig.\,\ref{fig2}(b) the resulting phase coherence lengths obtained from three devices are plotted as a function of $V_{g}$. For all samples minimal values of $1$~\textmu m to $2$~\textmu m are found for $l_{\varphi}$, while maximal values reach $5$~\textmu m at gate voltages around $V_{g}=1$~V. At $V_{g}=1$~V the Fermi level $E_F$ is in the bulk gap, as extracted from independent measurements for macroscopic Hall bars made from the the same material 
\cite{KozlovKvonOlshanetskyEtAl2014,KozlovBauerZieglerEtAl2016}. 
For $E_{F}$ in the gap the phase coherence lengths of the topological surface states are expected to be largest as backscattering is reduced and scattering into bulk states is suppressed.

Figure~\ref{fig3} shows $V_g$-dependent resistance traces of device w1 for a number of magnetic fields $B$ normalized to their value at $B=0$. The maximum of the longitudinal resistance, which comes out more clearly with higher magnetic fields, is usually ascribed to the charge neutrality point (CNP). In strained HgTe it is located slightly in the valence band \cite{KozlovKvonOlshanetskyEtAl2014}. Hence the valence band edge is located on the right hand side of the maximum. The $V_{g}$ locations of the conduction and valence band edges, indicated by arrows in Fig.\,\ref{fig3}, have been obtained by comparing to the results described by Kozlov \textit{et al.} (2014)\cite{KozlovKvonOlshanetskyEtAl2014,KozlovBauerZieglerEtAl2016}.

%Figure~\ref{fig3} compares $\rho_{xx}(V_{g})$ measured from a macroscopic reference Hall bar of the same wafer with the $V_{g}$-dependence of the resistance $R$ of device w1. The maximum of $\rho_{xx}(V_{g})$ is ascribed to the charge neutrality point. In strained HgTe it is located slightly in the valence band \cite{KozlovKvonOlshanetskyEtAl2014}. Hence the valence band edge is located on the right hand side of the maximum. The $V_{g}$ location of the conduction and valence band edge, indicated by dashed lines in Fig.\,\ref{fig3}, has been obtained from the reference sample using the analysis described in \cite{KozlovKvonOlshanetskyEtAl2014,KozlovBauerZieglerEtAl2016}. As the $R(V_{g})$ traces have a similar shape we assume in the following that the gap of the wires is located, at least approximately, in the same gate voltage range as for the bulk.

\begin{figure*}
	\includegraphics[width=2\columnwidth]{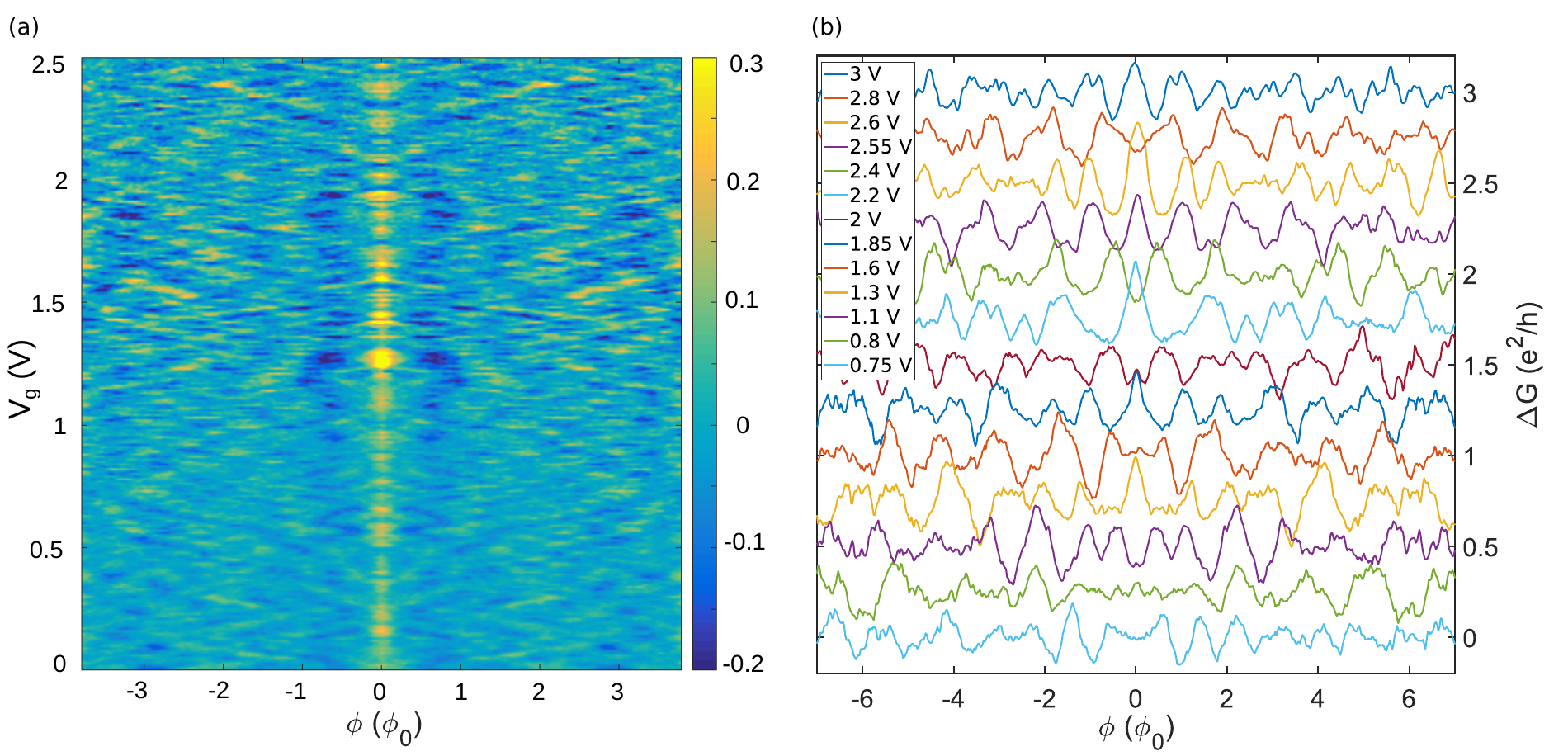} \caption{\label{fig4}
		Quantum conductance corrections of transport measurements in TI nanowires.
		(a) Color plot of conductance correction $\Delta G$ as a function of magnetic flux $\phi$ (in units of flux quantum $\phi_{0}$) and gate voltage $V_{g}$ from device w1. 
		(b) Selection of representative magnetoconductance curves from device w2 comprising a larger flux range. The different traces (offset for clarity) exhibit clear $\phi_{0}$-periodic behavior, as well as a switching of the phase with varying $V_{g}$.
	}
\end{figure*}

\begin{figure}
\includegraphics[width=1\columnwidth]{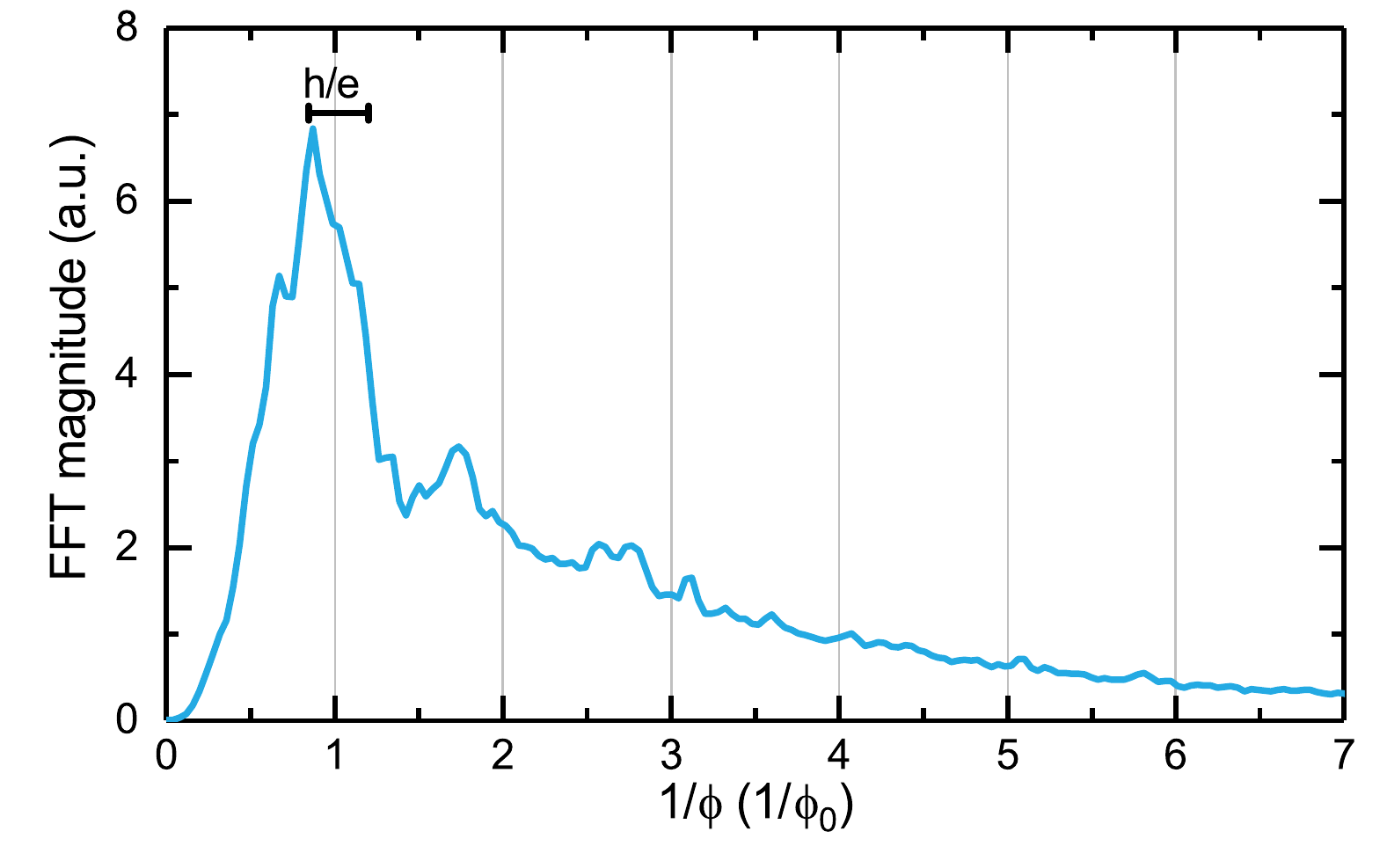} \caption{\label{fig5}
Fourier spectrum of nanowire magnetoconductance oscillations. 
Fast Fourier transform, calculated from the magnetoconductance of device w1 and averaged over all $V_{g}$ values, shows a dominant peak close to $1/\phi_0$, reflecting $\phi_{0}=h/e$-periodic oscillations. The bar at the peak denotes the expected range of the peak to occur, see text.
}
\end{figure}

Next we study in more detail the Aharonov-Bohm-type oscillations and in particular their $V_{g}$-dependence, here representatively discussed for device w1. To this end, we remove the small hysteresis of the superconducting magnet from the conductance and subtract a smoothly varying background $G_\mathrm{sm}(B)$ using a Savitzky-Golay filter \cite{Savitzky1964}. The filter was applied in a way that frequencies smaller than $1/2\phi_{0}$ are cut off. The resulting conductance $\Delta G(V_g;\phi)$ is shown in the color plot in Fig.\,\ref{fig4}(a). Here, the gate voltage spacing of neighboring traces is $\Delta V_{g}=0.01$~V. Corresponding data, but now as line cuts taken at different values of $V_g$ are shown in Fig.\,\ref{fig4}(b) for sample w2. $\Delta G$ shows dominant $\phi_{0}$-periodic oscillations over a large gate voltage range, as well as a pronounced phase switching between minima and maxima, {\em i.e.} additional conductance oscillations upon varying $V_g$ at fixed flux (to be analyzed in Sec.~\ref{subsec:subbands}). The $\phi_{0}$-periodicity is confirmed by a fast Fourier transform (FFT) analysis of traces covering a magnetic field range corresponding to 20$\phi_{0}$ in the case of device w1. As the period of the oscillations should be independent of the gate voltage  an average of the FFTs was taken in the voltage range $0$~to~$3$~V. The $V_{g}$-averaged FFT is shown in Fig.\,\ref{fig5} as a function of $1/\phi$. Here, we used the square geometry with a median width of $310$~nm for w1 and assumed that the topological surface states are $5$~nm below the surface to compute the magnetic flux $\phi=AB$ from $B$. The resulting FFT peak is located at $0.87/\phi_{0}$. 
The peak falls into the expected region of an $h/e$ signal, where the lower bound is given by the geometrical dimensions of the nanowire cross section as listed in Table\,\ref{tab1}. The upper bound is calculated for the case where the wave function of the surface states lies 8~nm within the TI bulk. 

To conclude, the distinct peak of the FFT close to $1/\phi_0$ implies that, at low temperatures,  transport is 
mediated by states extending phase-coherently across the entire surface of the weakly disordered nanowire. 
% However, this does not yet imply the topological character of these surface states that we will address in the following.

\subsection{
Subband-induced conductance modulation         
\label{subsec:subbands}
}
\begin{figure}
\includegraphics[width=1\columnwidth]{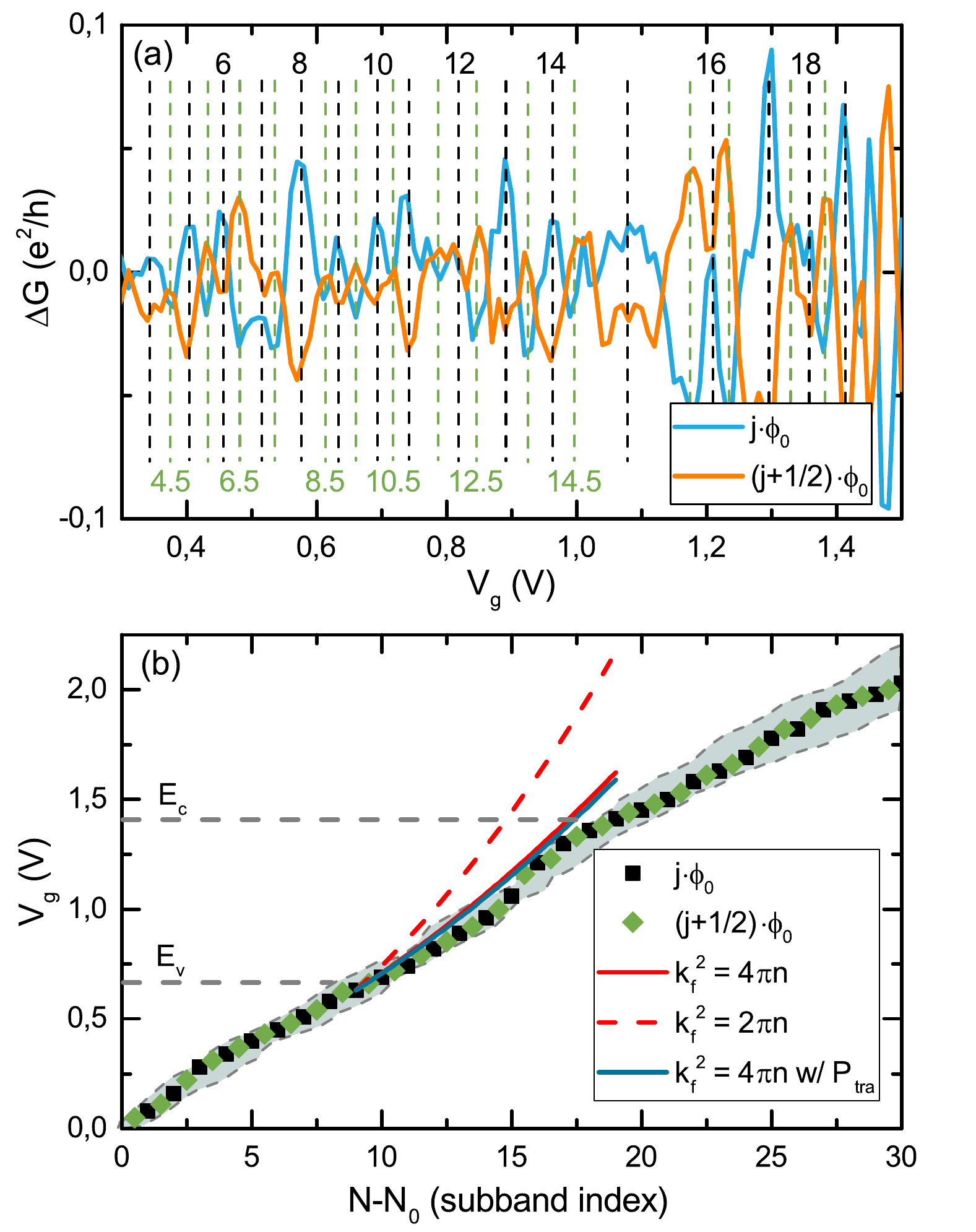} \caption{\label{fig6} 
(a) Conductance oscillations due to subband quantization. 
Conductance corrections $\Delta G$ as a function of gate voltage for integer (blue curve) and half-integer (orange curve) values of $\phi/\phi_{0}$, obtained by averaging several line-cuts taken along constant $\phi$ from the data set of device w1 in Fig.~\ref{fig4}(a). The curves exhibit a clear anti-correlated behavior. Each minimum-maximum pair (marked by vertical dashed lines) is assigned an index $N$ that can be associated with a subband of the wire, see text.
(b) Gate voltage position of conductance maxima from (a) as a function of (subband) index $N$ (with arbitrary offset $N_0$). Within the bulk gap (marked by dashed horizontal lines) the red curve segments denote expected behavior for $V_g(N)$ based on models for helical Dirac-type (solid line) and trivial spin-degenerate (dashed) surface states, see main text. 
Agreement of the conductance data with the former model implies the topological character of the surface states. For comparison, the blue line shows corresponding results for a Dirac surface state model, based on a trapezoidal cross section instead of a simple rectangular cross section. The grey area around the data points is obtained repeating the analysis while introducing an artificial error of up to $\pm0.2$ in $\phi_0$, showing that the method is robust against errors in $\phi_0$.
}
\end{figure}

The detection of the Aharonov-Bohm oscillations that switch their phase as a function of $V_{g}$ is by itself insufficient to confirm the topological nature of the underlying surface states. Therefore we have to go beyond previous analysis and aim at a quantitative description of the $\Delta G (V_g)$-oscillations,
which directly reflect the subband structure of the quasi-1D wires. 
$\Delta G (V_g;\phi)$ is extracted from line cuts taken at half integer and integer values of $\phi/\phi_{0}$ in
Fig.~\ref{fig4}(a) and plotted as blue, respectively orange, curves in Fig.~\ref{fig6}(a). To suppress the influence of
aperiodic conductance fluctuations we averaged over all positive and negative integer multiples of $\phi_{0}$ in the
dataset, i.e. $\pm\phi_{0}$,$\pm2\phi_{0}$ and $\pm3\phi_{0}$ to generate the orange curve in the case of the
representative device w1. Accordingly, the blue curve results from an average over positive and negative half-integer
multiples calculated from $\pm0.5\phi_{0}$, $\pm1.5\phi_{0}$ and $\pm2.5\phi_{0}$. The line cut for $\phi=0$ was omitted
due a large weak antilocalization-like background profile that renders background removal difficult. In Fig.~\ref{fig6}(a) the two conductance curves $\Delta G(V_g)$ obtained in this manner exhibit antiphase oscillations as a function of $V_g$, {\em i.e.} maxima of $\Delta G(V_g)$ for integer multiples of $\phi_0$ go along with minima of $\Delta G(V_g)$ for half integer multiples. 

In Secs.~\ref{sec:electrostatic model}-\ref{sec:conductance_simulations} we carry out an in-depth analysis associating these characteristics with the quantized subbands of helical surface states. 
Based on this study the analysis of the conductance oscillations in Fig.~\ref{fig6}(a) proceeds as follows: The minima in $\Delta G(V_g)$ correspond to a Fermi level at the edge of a subband since the corresponding high density of states (the van-Hove singularity) causes enhanced scattering. Accordingly, the gate voltage distance $\Delta V_g$ between two minima, corresponding to the period of the conductance oscillation, can be directly mapped onto the subband spacing. One key property that distinguishes topological surface states from their trivial counterparts is their degree of spin-degeneracy.  It directly affects $\Delta V_g$ since a spin-degeneracy of two, as for trivial surface states, implies that twice as many states need be filled compared to helical edge states.
This allows us to extract the spin degeneracy of the subbands involved by quantitatively analyzing the distance $\Delta V_g$ between adjacent conductance minima.

To increase the accuracy of the extracted $\Delta V_g$, we use the anticorrelation between the two curves for $j\phi_{0}$
and $(j+0.5)\phi_{0}$ with $j=\pm1,\pm2,...$ and the corresponding minima-maxima pairs in order to label the pairs by a running index and to obtain $\Delta V_g$. Pairs with the maximum corresponding to integer/half-integer flux quantum are labeled by an integer/half-integer index by black and green dashed lines, respectively. Since in strained HgTe the Dirac point is located in the valence band, we do not know the precise number of filled 1D subbands at a given gate voltage.
We thus start counting by fixing a reference voltage $V_0$ at which a certain (unknown) number $N_0$
of subbands is filled, and use the relative index $N-N_{0}$, with $N$ the total number of occupied subbands. In Fig.\,\ref{fig6}(b) the gate voltage at which a particular minimum-maximum pair occurs is plotted as a function of the corresponding index $N-N_0$. 

The connection between $\Delta G(V_g;\phi)$ and the nanowire bandstructure is {\em a priori} 
not obvious, since the nanowire surface charge carrier density, and ergo its capacitance, is inhomogeneous due to asymmetric gating. However, as we will detail in Secs.~\ref{sec:electrostatic model}-\ref{sec:conductance_simulations}, an analysis in terms of a {\it single} effective charge carrier density $n_\mathrm{eff}$, {\em i.~e.}~effective capacitance $C_{\rm eff}$, turns out to be fully justified. They are related to each other via
\begin{equation}
(n_\mathrm{eff}-n_{0})e=C_\mathrm{eff}(V_{g}-V_{0}), 
\label{eq:neff}
\end{equation}
where $V_{0}$ is the gate voltage at which the oscillation counting starts and $n_{0}$ the corresponding carrier density. 
For a 2D system the Fermi wave vector $k_{F}$ is given by 
\begin{equation}
k_{F}=\sqrt{(4\pi/g_s) n_\mathrm{eff}}
\label{eq:kF}
\end{equation}
with the spin degeneracy factor $g_s=1$ and $ g_s=2$ for spin-resolved topological surface states and spin-degenerate trivial surface states, respectively. Assuming a constant subband splitting $E/(\hbar v_F) = \Delta k_l=2\pi/P$ (in view of Eq.~(\ref{eq1})), the Fermi wave vector at a conductance minimum (subband opening) can be written as 
\begin{equation}
k_{F}=k_{0}+(N-N_{0})\Delta k_l \quad \mbox{with} \quad k_0=\sqrt{(4\pi/g_s) n_0} \, .
\label{eq:kF2}
\end{equation}
Combining Eqs.~\eqref{eq:neff}, \eqref{eq:kF} and \eqref{eq:kF2} yields the relation
\begin{equation}
V_{g}-V_{0}=\frac{g_s e}{4\pi C_\mathrm{eff}}\left[2k_{0}(N-N_{0})\Delta k_l+(N-N_{0})^{2}\Delta k_l^{2}\right]
\label{eq3}
\end{equation}
between gate voltage and subband level index $N$. The distance between two conductance minima depends essentially on the effective capacitance $C_\mathrm{eff}$ and wire circumference $P$.

Equation~(\ref{eq3}) is only valid for $V_{g}$ values for which $E_{F}$ is located
in the bulk gap. We estimate this $V_g$-range from the gate-voltage dependent resistance curves plotted in Fig.~\ref{fig3}. 
The corresponding region is highlighted by the dashed horizontal lines in Fig.\,\ref{fig6}(b) and matches the region of 
longest phase-coherence lengths $l_{\phi}$ discussed above.

The key quantity which is missing for comparing Eq.\,(\ref{eq3}) with the experimental data of Fig.\,\ref{fig6}(b) is the effective capacitance $C_\mathrm{eff}$, which is not directly accessible from experiment. 
Therefore, we resort to a numerical solution of the Poisson equation (described in Sec.~\ref{sec:electrostatic model}) 
and the definition of $C_\mathrm{eff}$ given in Sec.~\ref{subsec:full capacitance model}. 
The anticipated result of such a calculation is $C_{\mathrm{eff}}=3.987\cdot10^{-4}$~F/m$^{2}$, 
carried out for the geometry and dielectric constants of the present wire. 
The curves obtained from Eq.~(\ref{eq3}) both for spin helical and trivial surface states
are shown in Fig.\,\ref{fig6}(b). In the gap region the experimental data points are best described by the model invoking 
spin-helical surface
states (solid lines); trivial surface states lead to a steeper slope (red dashed line) and fail to describe the experiment. It would be interesting to apply this kind of analysis to wires with trivial surface states\cite{Guel2014}.

\begin{figure}
\includegraphics[width=1\columnwidth]{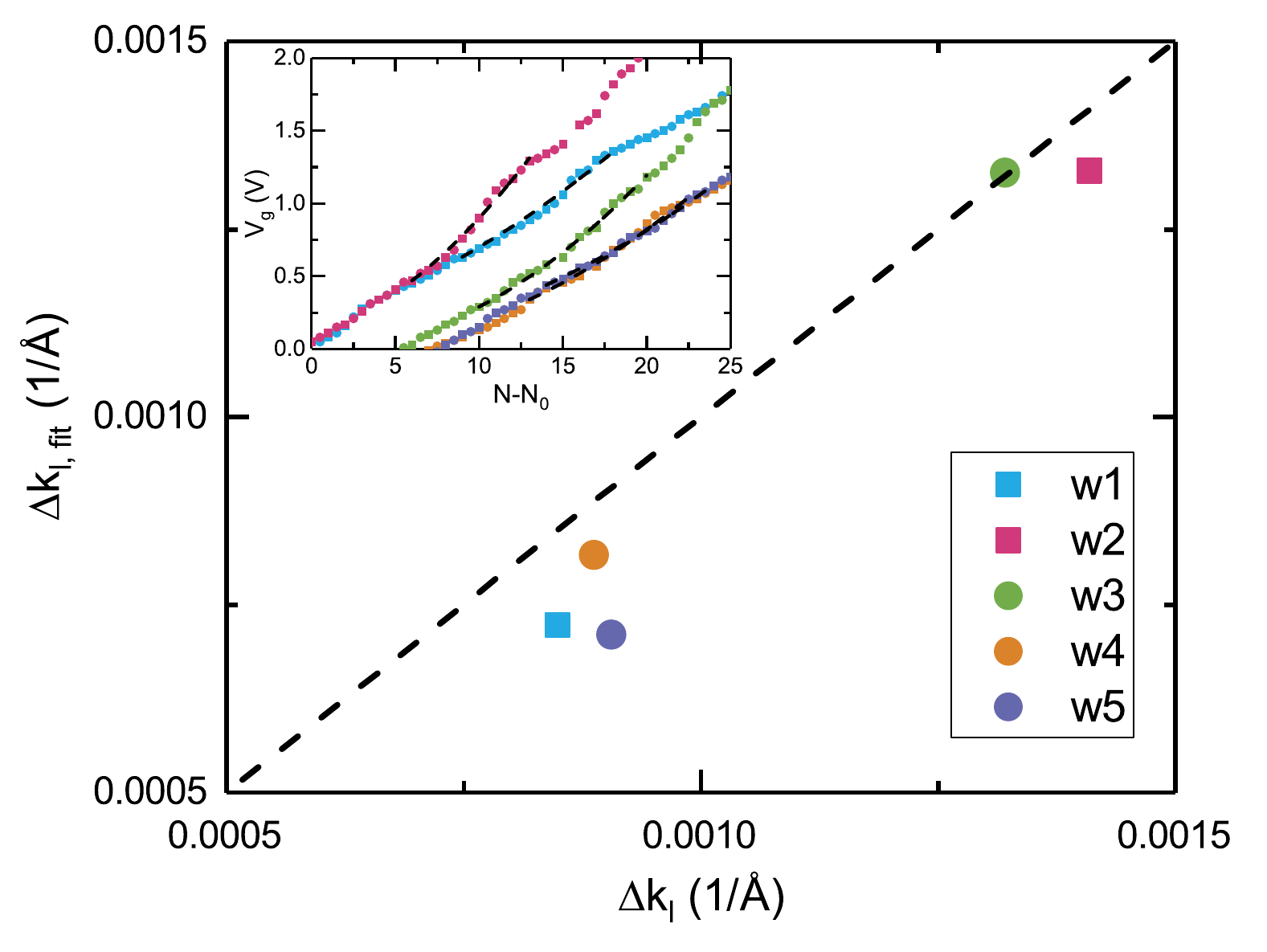} \caption{\label{fig7} 
Analysis of subband spacing. The inset shows the dependencies $V_{g}(N-N_{0})$  of samples w1-w5. The black dashed lines are best fits (within the bulk band gap) based on Eq.~(\ref{eq3}) with $\Delta k_{l}$ as the only fit parameter. Main panel: Best fitting values $\Delta k_{l,\mathrm{fit}}$ of samples w1-w5 are plotted versus $\Delta k_{l}=2\pi/P$, determined from the nanowires' circumference $P$ taken from SEM micrographs. The dashed straight line marks the condition $\Delta k_{l}=\Delta k_{l,\mathrm{fit}}$. Devices whose data is presented in this article are marked by squares.}
\end{figure}

The analysis described in this section was carried out for five devices with different cross sections $A$. For each device the nanowire circumference $P$, which defines the subband splitting $\Delta k_{l}=2\pi/P$, was extracted from the SEM micrographs. Alternatively, one can extract the circumference by fitting the data points, shown for samples w1 - w5 in the inset of Fig.~\ref{fig7}, with Eq.~(\ref{eq3}). Using the corresponding effective capacitances $C_\mathrm{eff}$, the only remaining fit parameter is $\Delta k_{l,\mathrm{fit}}$. The best fits in the relevant $V_g$-ranges for all five investigated samples are shown in the inset of Fig.~\ref{fig7}. In the main panel of Fig.~\ref{fig7} the extracted values of $\Delta k_{l,\mathrm{fit}}$ are plotted versus $\Delta k_{l}$. The plot shows that $\Delta k_{l}\approx\Delta k_{l,\mathrm{fit}}$ within experimental accuracy,
thus confirming the suitability of the analysis.

In the following we present an in-depth study of the nanowire electrostatics, band structure and magnetoconductance,
that provides the theoretical basis for the analysis of Fig.~\ref{fig6} indicating the existence of
helical surfaces states.

\section{Electrostatics of wire geometry  \label{sec:electrostatic model}}

Since the capacitance used in the preceding section is not experimentally accessible we need to resort to numerical analysis. Furthermore, in the experimental setup, the Fermi energy is locally tuned by a gate electrode that covers roughly the upper part of the 
nanowire [see Fig.~\ref{fig1}(c)] and induces a non-uniform surface electron density $n(z,s)$, where $z$ is the longitudinal
($0 < z < L$) and $s$ the circumferential coordinate ($0 < s < P=2w+2h$). To account for the inhomogeneous 
charge density in our subsequent transport simulations while keeping the model as simple as possible, 
we assume that $n(z,s)=n(s)$ is constant along the $z$-direction. We implemented the finite-element based partial-differential 
equation (PDE) solver \mbox{\textsc{FEniCS}} \cite{FEniCS} combined with the mesh generator \textsc{gmsh} \cite{gmsh} to obtain 
the gate capacitance, considering a 2D electrostatic model for the geometry sketched in Fig.~\ref{fig1}(c). 
Furthermore, the HgTe wire along with the Au top gate are both assumed perfectly metallic with vanishing electric field 
in the interior, which implies Dirichlet-type boundary conditions on the corresponding surfaces, {\em i.e.}
the electric potential $u(x,y)=V_g$ on the boundary of the Au top gate and $u(x,y)=0$ on the HgTe nanowire.

\begin{figure}
	\includegraphics[width=0.96\columnwidth]{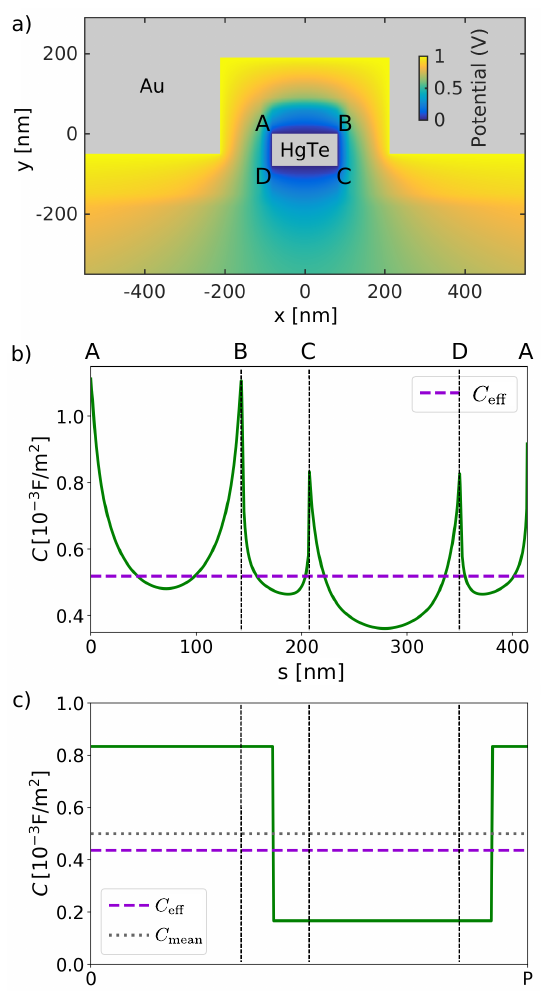}	
	\caption{\label{fig:Ming-Hao_Gate} Electrostatics of gated nanowires. (a) 
Electric potential $u(x,y)$ across the heterostructure of device w2. (b) Capacitance profile $C(s)$ along the circumference of 
nanowire w2 obtained via Eq.~(\ref{eq:electric}) from the electric potential depicted in (a). 
(c) Capacitance profile for the simplified model (see text)  for parameters $P=460$ nm and 
$C_\textrm{mean}=\num{5d-4}\si{\farad\per\square\metre}$.}
\end{figure}

The PDE solver numerically yields solutions of the Laplace equation $\nabla^2 u(x,y)=0$ for the heterostructure,
an example with $V_g=1\mathrm{V}$ is given in Fig.~\ref{fig:Ming-Hao_Gate}(a). 
The induced surface charge density $n(s)$ is then given by the gradient of the electric potential at the nanowire surface
according to \cite{Liu2013}
\begin{equation} 
n(s)=(\epsilon_r\epsilon_0/e) (\nabla u) \cdot \hat{e}_n \, ,
\label{eq:electric}
\end{equation}
where $\hat{e}_n$ is the unit vector defining the surface normal, $\epsilon_r$ is the dielectric constant of the insulating 
layer surrounding the nanowire ($\epsilon_r=3.5$ for SiO$_2$ and $\epsilon_r=13.0$ for HgCdTe) and 
$\epsilon_0$ is the vacuum permittivity. 
The gate capacitance per unit charge, $C(s)/e$, defined as the surface electron density $n(s)$ per gate voltage $V_g$,
is then exported from the PDE solver. The surface charge density at arbitrary gate voltages is obtained by exploiting 
the linearity $n(s)= [C(s) / e] V_g$ without the need for repeating the electrostatic simulation.

Figure \ref{fig:Ming-Hao_Gate}(b) shows the capacitance $C(s)$ resulting from the simulation. 
The large spikes in the capacitance stem from the sharp edges of the HgTe nanowire. Electrostatic simulations for a trapezoidal wire cross-section with smoother gate profiles (not shown) show that our simple model overestimates the effective capacitance by $ <4\% $. Assuming metallic surface states and ignoring thus the quantum capacitance also leads to capacitance values which are $ ~5\% $ larger. The finite length of our wires which is ignored in our 2D electrostatic model underestimates the capacitance by the fringe fields at the ends. By analogy with finite length cylindrical capacitors the error for our wires is estimated to be $ \leq10\% $ \cite{Das1997}. As the errors due to the idealizations within our model tend to compensate each other, we expect the calculated values to be quite accurate.

\section{Bandstructure of gated nanowire}
\subsection{Dirac surface Hamiltonian \label{sec:Dirac surface Hamiltonian}}

Assuming that the bulk is insulating we model the HgTe nanowire by means of the Dirac surface Hamiltonian 
\begin{equation}
	H = v_F [p_z \sigma_z + (p_s + e A_s) \sigma_y] \, ,
	\label{eq:2d_dirac}
\end{equation}
where $A_s$ is the vector potential which creates the longitudinal magnetic field.   
The numerical results presented in the following were obtained using the python software package Kwant \cite{Groth2014}. 
Since Kwant utilizes tight-binding models, Eq.~(\ref{eq:2d_dirac}) needs to be discretized leading to Fermion 
doubling \cite{Susskind1977, Stacey1982}. In order to circumvent this, we add a small term quadratic in the momentum 
to the Hamiltonian which removes the artificial valleys from the considered energy range (for a recent discussion 
see \cite{Habib2016}).

\begin{figure*}
	\includegraphics[width=2\columnwidth]{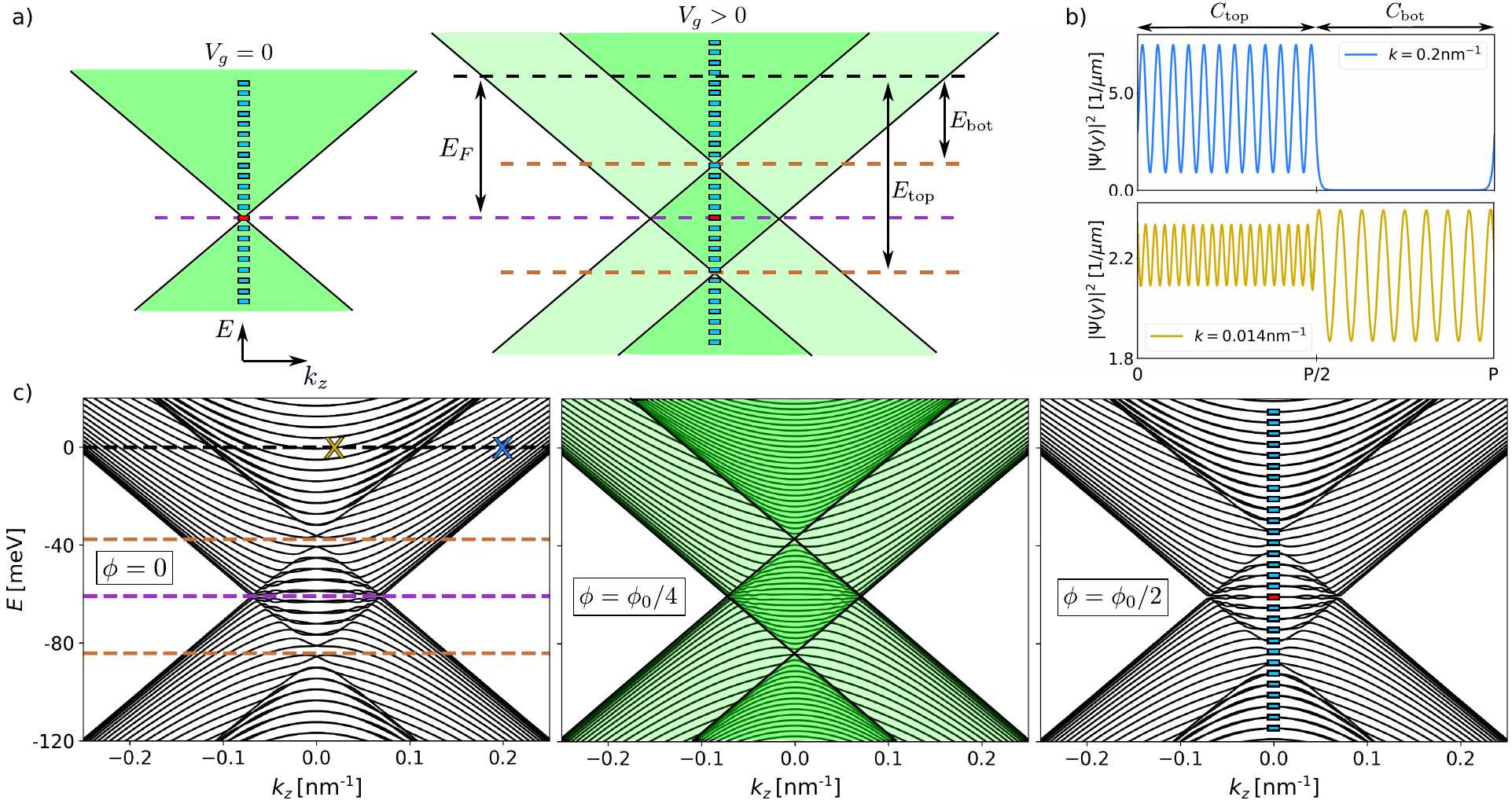}
	\caption{\label{fig:simple_step_cap} 
Gate-dependent splitting of Dirac-type nanowire bandstructure for simplified step capacitance model for
top and bottom surface [see Fig.~\ref{fig:Ming-Hao_Gate}(c)].
(a) Sketch of the bandstructure for zero (left) and finite (right) gate voltage. 
Subband bottoms at $k_z=0$ are indicated by blue rectangular boxes. 
 (b) Probability distributions of two representative states along wire circumference with capacitances $C_{\rm top}$ and
$C_{\rm bot}$ at top and bottom surface. The state in the upper (lower) panel corresponds to an energy marked by a blue (yellow) cross in panel (c).
  (c) Calculated bandstructures for three magnetic fluxes $\phi/\phi_0 = 0, 1/4, 1/2$
for $V_g = 1 V$ and $v_F = \SI[per-mode=symbol]{5e5}{\metre\per\second}$.
In the left panel the dashed black horizontal line marks the Fermi level and the dashed yellow lines positions of shifted Dirac
points..
 In the middle panel areas corresponding to flux-sensitive (insensitive) energy levels are marked by  dark (light) green color.}
\end{figure*}

The magnetic flux and the curvature-induced Berry phase are implemented via the boundary condition 
\begin{equation}
	\Psi(z, s + P) = \Psi(z, s) \mathrm{e}^{i 2 \pi(\phi/ \phi_0 + 1/2)} \, .
\end{equation}
We further account for the effect of the top gate by adding the onsite energy
(see corresponding Eq.~(\ref{eq:kF}) with $g_s=1$)
\begin{align}
	E_{\textrm{gate}}(s) = -\hbar v_F \sqrt{4 \pi V_\textrm{g} C(s) / e} \, ,
	\label{eq:gate_potential}
\end{align}
which induces the correct charge density $n(s)$ along the nanowire circumference.
\subsection{Gate effect: Simplified capacitance model 
	\label{sec:simplified capacitance model}}

In order to illustrate the effect of the gate-induced potential on the bandstructure of a nanowire, 
we start with a simple step-shaped capacitance [see Fig.~\ref{fig:Ming-Hao_Gate}(c)] before examining the more 
realistic case shown in Fig.~\ref{fig:Ming-Hao_Gate}(b). In this simplified model, $C(s)$ is determined by two 
capacitance values, one for the top surface $C_{\textrm{top}}$ and one for the bottom surface $C_{\textrm{bot}}$,
neglecting separate profiles at the narrow side surfaces. 
We choose $C_{\textrm{top}} / C_{\textrm{bot}} = 5$ for didactic purposes; in the experiments the ratio is $ \approx 2$.

In the following, we first use a sketch of the resulting bandstructure in Fig.~\ref{fig:simple_step_cap}(a) to explain the 
mechanisms that lead to the corresponding numerical results shown in Fig.~\ref{fig:simple_step_cap}(c).
For $V_g = 0$ the bandstructure is given by a simple 1D Dirac cone [left panel in Fig.~\ref{fig:simple_step_cap}(a)] 
with quantized subbands owing to the finite circumference. The positions of the subband minima at $k_z=0$ are marked by 
blue rectangular boxes. The flux through the nanowire is chosen to be $0.5 \phi_0$ implying a state at zero energy 
(marked by a red box).
For $V_g  > 0$  the Dirac cone splits, so that the distances between the common Fermi level $E_F$ and the two Dirac points are
$E_{\textrm{top}} = \hbar v_F \sqrt{4 \pi n_{\textrm{top}}}$ and $E_{\textrm{bot}} =\hbar v_F \sqrt{4 \pi n_{\textrm{bot}}}$ 
[right panel in Fig.~\ref{fig:simple_step_cap}(a)].  The difference arises since the top surface is filled faster than 
the bottom one.  The Fermi energy is given by the average $E_F = (E_\textrm{top} + E_\textrm{bot}) / 2 = \hbar v_F \sqrt{\pi
	V_g / e} \left(\sqrt{C_\textrm{top}} + \sqrt{C_\textrm{bot}} \right)$.

Interestingly, the splitting of the Dirac cone does not influence the $k_z=0$ subband spacing, which is perfectly preserved.
Furthermore, for $k_z \neq 0$ the states can be divided into two groups. States with energies in the dark green regions in 
Fig.~\ref{fig:simple_step_cap} extend over the entire wire circumference and consequently are flux-sensitive, whereas 
states with energies in the light green regions are localized on the upper or lower surface and hence are
not susceptible to flux changes. Representative examples of both kinds of states are shown in Fig.~\ref{fig:simple_step_cap}(b). 
The reason for this peculiar behavior is Klein tunneling \cite{Allain2011}.  Modes at $k_z=0$ perpendicularly hit the potential step associated with the capacitance profile of Fig.~\ref{fig:Ming-Hao_Gate}(c)
and are thus unaffected by its presence -- Klein tunneling is perfect.  On the other hand, for $k_z\neq0$ the light green regions host states from one Dirac cone only,
the dark green ones from both.  Since $k_z$ is conserved during tunneling, 
it is only in the dark green regions that electrons can Klein-tunnel from one cone to the other, yielding hybridized extended states. 

The different flux sensitivity of the two classes of states is also reflected in the numerical bandstructures shown in Fig.~\ref{fig:simple_step_cap}(c) for three different fluxes: While the hyperbolic energy levels in the regimes
corresponding to the light green areas are identical (on scales resolved in the figures) for all three fluxes, the energy 
levels belonging to the dark green areas obviously change with varying flux. We note in passing that trivial surface states which might form at the etched side surfaces do not contribute to the oscillations and thus to our analysis as they are localized at the sides.

\subsection{Gate effect: realistic capacitance model 
	\label{subsec:full capacitance model}}

\begin{figure*}
	\includegraphics[width=2\columnwidth]{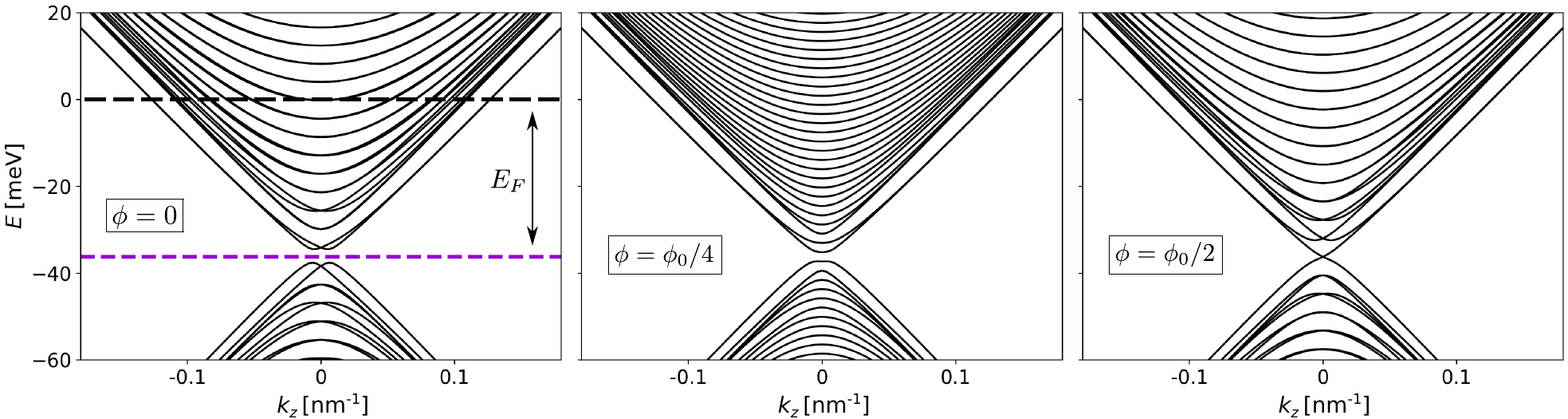}	
	\caption{\label{fig:full_cap_model} Calculated bandstructures of device w2 based on the realistic capacitance model 
		[see Fig.~\ref{fig:Ming-Hao_Gate}(b)] for $V_g=0.3 \mathrm{V}$ and three fluxes  $\phi/\phi_0 = 0, 1/4, 1/2$ (from left to right). 
		The spacing between subband minima at $k_z=0$ is constant for given flux.}
\end{figure*}

We now use these insights to examine the results obtained with the more realistic capacitance profile shown in 
Fig.~\ref{fig:Ming-Hao_Gate}(b). We present results for the nanowire geometry of sample w2;
the other geometries do not qualitatively change the results.
The corresponding bandstructures are presented in Fig.~\ref{fig:full_cap_model} for three different
magnetic fluxes and $V_g=0.3$ V.
The interpretation of the bandstructure is not as simple as in the step capacitance model, as there are now four
nanowire surfaces involved, each with a more complicated capacitance profile.
Furthermore, the difference between the ``potential bottoms'' on each surface is smaller.  
However, the main features discussed in Sec.~\ref{sec:simplified capacitance model} 
are still present, since they are actually independent of the capacitance profile shape.

Notably, Fig.~\ref{fig:full_cap_model} shows that the $k_z=0$ subband spacing is perfectly preserved, 
even though the transversal wave vector $k_l (s) = \sqrt{4 \pi n(s)}$ is now $s$-dependent.
Indeed, we can generalize the discussion of the simplified step capacitance model:
For $k_z=0$ the motion is purely angular, and thus Klein tunneling is perfect irrespective of the complexities
of the potential profile.  Hence, the electron wave function experiences
an average of the gate potential, whose corresponding wave vector average fulfils
\begin{align}
\left< k_l(s) \right> \equiv \frac{1}{P}	\int\limits_{0}^{P} \textrm{d}s \, k_l(s) = \frac{2 \pi}{P}(l + 0.5 - \phi
/ \phi_0) \, .
\end{align}
This leads to a gate-independent subband spacing $\Delta k_l = 2 \pi / P$ that can be also derived by solving the 
Dirac equation for $k_z=0$.
Moreover, in the experimentally relevant parameter range, the Fermi level stays always above both Dirac points for 
$V_g >0$, and the subband minima at $E_F$ are always at $k_z=0$.

For a general capacitance profile $C(s)$ the Fermi energy is given by the average
\begin{equation}
E_F = \hbar v_F \sqrt{4 \pi V_g/e} \left< \sqrt{C(s)} \right> = \hbar v_F \sqrt{4 \pi V_g  C_\textrm{eff} /e},
\label{eq:E_CM}
\end{equation}
where the effective value $C_\textrm{eff} \equiv \left< \sqrt{C(s)} \right>^2$ enters instead of the
mean $C_\mathrm{mean} \equiv \left < C(s) \right>$ with $\left<\ldots \right>$ denoting
the mean value along the circumference.
The value for  $C_\textrm{eff}$ calculated from the capacitance profile of w1
was used in the analysis of the experimental data in Fig.~\ref{fig6}.
Note that the difference between $C_\mathrm{eff}$ and $C_\mathrm{mean}$
is given by the variance of the capacitance profile $\mathrm{Var}(\sqrt{C(s)}) = C_\mathrm{mean} - C_\mathrm{eff}$. 
For the step capacitance model with $C_\mathrm{top} = 5 \, C_\mathrm{bot}$ this difference is clearly 
visible [$C_\mathrm{eff} / C_\mathrm{mean} \approx 0.87$, see Fig.~\ref{fig:Ming-Hao_Gate}(c)] but in the experiments 
it is typically quite small (e.g. for w2  $C_\mathrm{eff} / C_\mathrm{mean} \approx 0.985$).

As will become evident in the following, the perfect subband quantization and the fact that
the subband minima are located at $k_z=0$ are crucial ingredients to probe the signature of the Dirac surface states in transport.

\section{Conductance simulations
\label{sec:conductance_simulations}}

The conductance simulations presented in the following were carried out by using an extended version of the tight-binding
model introduced in Sec.~\ref{sec:Dirac surface Hamiltonian}. The extended system includes coupling of the nanowire to leads
with the same geometry as the nanowire but with negative onsite energy (i.e. highly-doped leads) to account for the wide leads
in the experiment.
Residual disorder in the wires used in experiment is modelled by adding a random disorder potential $V(\boldsymbol{r})$
defined through the correlator
\begin{equation}
        \left< V(\boldsymbol{r}) V(\boldsymbol{r}') \right> = K_0 \frac{(\hbar v_F)^2}{2 \pi \xi^2} \mathrm{e}^{-|\boldsymbol{r} -
\boldsymbol{r}'|^2/2 \xi^2} \, .
\label{eq:disorder}
\end{equation}
Here, $\xi$ is the correlation length and $K_0$ determines the disorder strength {\footnote{For a detailed discussion of scattering in quasi-ballistic Dirac nanowires, see \cite{Dufouleur2017}}.

Figure~\ref{fig:eval_G} shows the conductance $G(V_g)$ for $\phi=0$ and $\phi=0.5 \phi_0$. 
Both curves exhibit distinct oscillations on top of an increasing conductance background (similar to corresponding calculations
in Ref.~\cite{BardarsonBrouwerMoore2010}). 
The anti-correlated behavior of the two oscillatory curves is due to the flux-sensitivity of the states near $k_z=0$. 
The oscillatory behavior can be explained by the bandstructure: 
Whenever the Fermi energy approaches the bottom of one of the disorder-broadened subbands, the high density of states (associated
with a van-Hove singularity) causes enhanced scattering and thus leads to a reduction of the conductance. 
By further increasing the gate voltage, the additional conductance channel fully opens and the Fermi energy leaves the vicinity of the 
van-Hove singularity, both effects leading to an increasing conductance. 
Thus, the conductance oscillation, related to successive subband opening, is a fingerprint of the bandstructure 
at $k_z=0$, as also observed in the transport measurements.
Due to the undisturbed subband quantization at $k_z=0$, the distance between two conductance minima (one oscillation period) 
corresponds to a change of $\Delta k_l = 2 \pi / P$ in the Fermi wave vector, as in the case without any inhomogeneity caused by the top gate. However, the inhomogeneity enters via $C_\mathrm{eff}$, which determines the Fermi wave vector 
$k_F = \sqrt{4 \pi V_g  C_\textrm{eff} /e}$ [see Eq.~(\ref{eq:E_CM})].
This justifies {\em a posteriori} the validity of Eq.~(\ref{eq3}) and the procedure used in 
Sec.~\ref{subsec:subbands} to analyze the experimental data if the surface states are Dirac-like.

\begin{figure}
	\includegraphics[width=1\columnwidth]{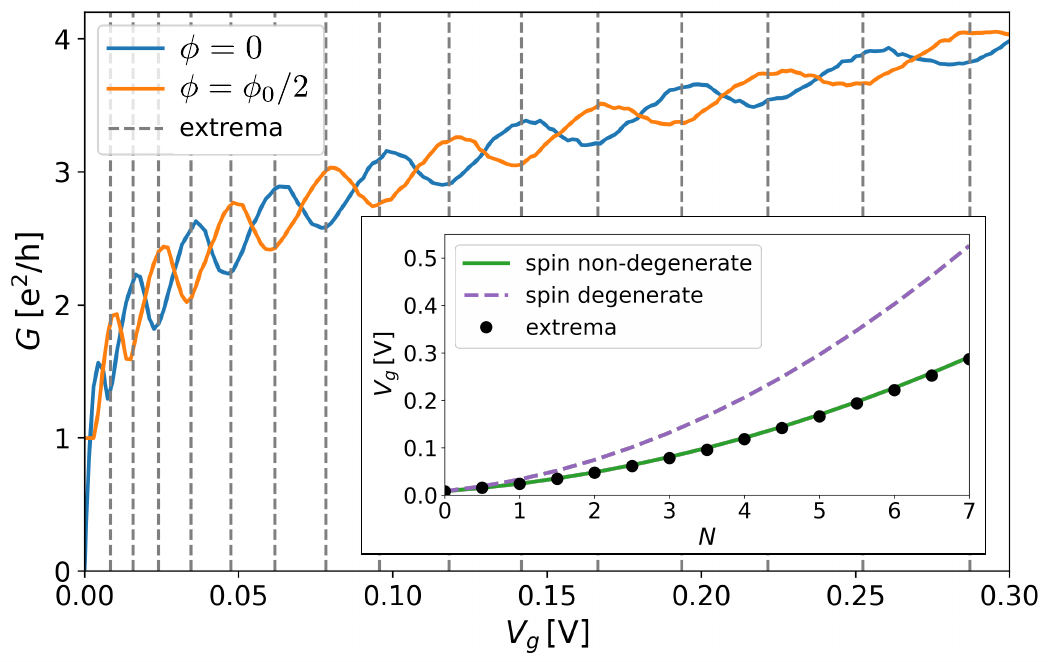} 
	\caption{\label{fig:eval_G}
		Calculated disorder-averaged conductance as a function of gate voltage (starting from the Dirac point)
		for zero flux (blue curve) and half a flux quantum (red curve). Calculations are performed for the nanowire
		geometry and using $C_\mathrm{eff}$ of sample w2.
		The average is taken over $10^3$ disorder configurations, based on the impurity model (Eq.~(\ref{eq:disorder})) with disorder
		strength $K_0 = 0.2$ and correlation length $\xi = P / 100$.
		Minima-maxima pairs in the conductance are marked with gray vertical lines.
		The inset shows the gate position of the minima-maxima pairs as a function of subband index $N$, following the same evaluation
		as performed in Fig.~\ref{fig5}(b) for the experimental data.}
\end{figure}

Trivial surface states, however, are expected to have a quadratic dispersion, {\em i.e.} Klein tunneling is absent. 
Thus, the question arises whether the $\Delta G(V_g)$-oscillations would still allow for the quantitative analysis performed
in Sec.~\ref{subsec:subbands}. Our simulations (not shown) for a quadratic dispersion reveal that for realistic gate-induced 
potentials the subband minima are still at $k_z=0$ but now states below $E_F$ can be confined and thus become flux-insensitive. 
Moreover, the subband quantization $\Delta k_l = 2 \pi / P$ is no longer preserved.
However, for $E_F$ in our experimental range the $k_z=0$ modes, which lead to the conductance oscillations,
have an angular motion energy that is much larger than the gate potential. 
These states are thus flux-sensitive and the subband quantization is only mildly affected (the degeneracy with respect 
to angular momentum is lifted). The lifting of the degeneracy is small compared to the subband spacing and cannot be observed 
within the experimental precision. For trivial surface states, we therefore indeed expect to measure a conductance 
with a similar shape as for Dirac states but with minima-maxima pairs following the spin-degenerate version ($g_s=2$) of 
Eq.~(\ref{eq3}).

The inset of Fig.~\ref{fig:eval_G} shows a similar evaluation of the minima-maxima pairs as was done for the
experimental data in Fig.~\ref{fig5}(b). 
The green parabolic curve describing spin-helical Dirac states according to Eq.~(\ref{eq3}) (with $g_s = 1$) 
matches perfectly with the results (black bullets) from the analysis of the gate-dependent conductance extrema, 
whereas the purple curve which holds for spin-degenerate trivial surface states ($g_s=2$) is way off.
Since we used Eq.~(\ref{eq:gate_potential}) (with $g_s=1$) to simulate the gate effect, the agreement was essentially expected. 
However, this analysis shows that if the surface states in 3D HgTe nanowires are Dirac-like, one should be able to 
obtain this signature of spin non-degenerate states by conductance measurements, despite the complicated potential 
profile induced by the top gate, as long as $C_\mathrm{eff}$ is known.

\section{Conclusions and outlook}

We fabricated nanowires based on strained HgTe and investigated in detail their peculiar transport 
properties in a joint experimental and theoretical effort. 
With regard to topological insulator properties, 
HgTe-based systems represent an interesting alternative 
to Bi-based systems, as surface states in the former appear to be well decoupled from bulk states, and additionally
feature high surface mobilities \cite{KozlovKvonOlshanetskyEtAl2014}. 
The nanowires were built out of strained bulk systems in a well controlled way.
In particular, we demonstrated that in these mesoscale conductors the topological properties of the 
corresponding bulk systems prevail, and appear in combination with quantum coherent effects: 
The observed $h/e$-periodic Aharonov-Bohm-type conductance modulations due to a coaxial flux
clearly indicate that transport along the wires is indeed both surface-mediated and quasi-ballistic, and additionally 
phase-coherent at micron scales. At low temperatures we moreover found that the extracted phase-coherence lengths 
are increased up to 5 $\mu$m upon tuning the wire Fermi energy into the bulk band gap, were topological surface transport
is singled out.

Besides the Aharonov-Bohm oscillations, we observed and examined in detail further distinct conductance
oscillations appearing as a function of a gate voltage. We showed that the spacing of the observed regular
gate-dependent oscillations reveals the topological nature of the surface states: The gate dependence
is only compatible with a model assuming non-degenerate (Dirac-type helical) surface states, and rules out usual spin-degenerate states. 
This identification required on the theory side a quantitative electrostatic calculation
of the gate-induced inhomogenous charge carrier density and associated capacitance of the whole (wire plus gate) system. 
The latter entered the evaluation of the transport data, and was furthermore integrated into
numerical tight-binding magneto-transport calculations to show that the gate-dependent conductance oscillations obtained indeed agree
with experiment. This theoretical analysis also shed light on the role of Klein tunneling for the
flux-sensitive surface states extending around the entire wire circumference, which govern the wire bandstructure.

Our finding that spin non-degenerate states exist on the wire surface suggests in particular that
for $\phi/\phi_{0}=0.5$, where time reversal symmetry is restored, the total number of left- or right-moving states 
is always odd at arbitrary Fermi level, implying a topologically protected perfectly transmitted mode.
In the presence of an $s$-wave superconductor, which opens a gap in the nanowire bandstructure via the
proximity effect, Majorana fermions are expected to appear at the endings of the 3DTI wire, if the latter hosts 
such an odd number of states at $E_{F}$ \cite{Cook2011,deJuan2014}. 
This opens up the interesting possibility of switching from a topologically trivial (even number of
states at $E_{F}$) to a non-trivial situation (odd number of states at $E_{F}$) 
by adding half a flux quantum through the wire's cross section. 
This notably led to a recent proposal to use proximitized TI nanowires as building blocks
for (coupled) topological Majorana qubits and networks of those \cite{Manousak2017}. 
While such proposals usually assume a uniform carrier density around the wire circumference, in experiments such density is 
expected to be strongly inhomogeneous, especially if the Fermi level is tuned via a top/back gate voltage. 
This raises the question of how substantially the topological behavior would be affected by strong inhomogeneities.
Here we demonstrated through the quantitative analysis of our data that the essential characteristics of topological transport
are indeed preserved under realistic experimental conditions.

To conclude, our work puts forward HgTe-based topological insulator nanowires as a promising and realistic platform 
for exploring a wealth of phenomena based on spin-momentum locked quantum transport and topological superconductivity.

%\section*{Acknowledgments}
\begin{acknowledgments}
This work was supported by {\em Deutsche Forschungsgemeinschaft} (within Priority Program SPP 1666 {\em ''Topological Insulators''})
and the {\em ENB Doktorandenkolleg ''Topological Insulators''}. Support by RF President Grant No. MK-3603.2017.2 and RFBR Grant No. 17-42-543336 is also acknowledged.
We thank J.~Bardarson, J.~Dufouleur, S.~Essert and E.~Xypakis for useful conversations.
\end{acknowledgments}

%\bibliographystyle{bibtex_nature_style}
%\bibliography{citations}

%

\end{document}